\documentclass[letterpaper, 10 pt, conference]{ieeeconf}  

\IEEEoverridecommandlockouts                              

\usepackage{amssymb,upgreek,nicefrac,amsmath}
\usepackage{sgame}
\usepackage[mathscr]{eucal}
\usepackage[dvips]{epsfig}
\usepackage[dvips]{graphicx}
\usepackage[section]{placeins}
\newcommand{\Cc}{C_b}

\newcommand{\cA}{{\mathcal{A}}}
\newcommand{\cX}{{\mathcal{X}}}
\newcommand{\cZ}{{\mathcal{Z}}}
\newcommand{\cS}{{\mathcal{S}}}

\newcommand{\Bor}{{\mathfrak{B}}}
\newcommand{\cI}{\mathcal{I}}
\newcommand{\df}{\doteq}
\newcommand{\reward}[2]{u_{#1}(#2)}
\newcommand{\Prob}{{\mathbb{P}}} 

\newcommand{\Simplex}[1]{\Delta\left(#1\right)}

\newcommand{\Dirac}[1]{\boldsymbol{\delta}_{#1}}

\newtheorem{definition}{Definition}[section]

\newtheorem{lemma}{Lemma}[section]
\newtheorem{theorem}{Theorem}[section]
\newtheorem{proposition}{Proposition}[section]
\newtheorem{property}{Property}[section]

\newcommand{\tr}{^{\mathrm T}}

\newcommand{\magn}[1]{\left\vert #1 \right\vert}

\newif\ifproofs
\proofstrue

\author{Georgios C. Chasparis\thanks{G.C. Chasparis is with the Department of Data Analysis Systems, Software Competence Center Hagenberg GmbH, Softwarepark 21, A-4232 Hagenberg, Austria, E-mail: georgios.chasparis@scch.at.}}

\title{Aspiration-based Perturbed Learning Automata\thanks{\ifproofs \else For the sake of brevity, all proofs have been omitted and can be found in \cite{chasparis_aspiration_2018}.\fi\newline
This work has been partially supported by the European Union grant EU H2020-ICT-2014-1 project RePhrase (No. 644235). It has also been partially supported by the Austrian Ministry for Transport, Innovation and Technology, the Federal Ministry of Science, Research and Economy, and the Province of Upper Austria in the frame of the COMET center SCCH.}}

\begin{document}

\maketitle
\thispagestyle{empty}
\pagestyle{empty}

\begin{abstract}
This paper introduces a novel payoff-based learning scheme for distributed optimization in repeatedly-played strategic-form games. Standard reinforcement-based learning schemes exhibit several limitations with respect to their asymptotic stability. For example, in two-player coordination games, payoff-dominant (or efficient) Nash equilibria may not be stochastically stable. In this work, we present an extension of perturbed learning automata, namely aspiration-based perturbed learning automata (APLA) that overcomes these limitations. We provide a stochastic stability analysis of APLA in multi-player coordination games. We further show that payoff-dominant Nash equilibria are the only stochastically stable states. 
\end{abstract}

\section{Introduction} \label{sec:Introduction}

Multi-agent formulations can be used to tackle distributed optimization problems due to their reduced communication and computational complexity. In such formulations, agents make their \emph{own} decisions repeatedly over time trying to maximize their \emph{own} utility/performance function. However, due to the interdependencies among agents' utility functions, \emph{local} (or \emph{distributed}) optimization does not necessarily imply \emph{global} (or \emph{centralized}) optimization. The problem becomes even more challenging when the utility function of each agent is unknown, and only measurements of this function (possibly corrupted by noise) are available. For this reason, there have been several efforts towards the design of \emph{distributed payoff-based learning dynamics} for convergence to globally optimal outcomes.

Naturally, several such distributed optimization problems can be formulated as strategic-form games. A rather common objective is then to derive conditions under which convergence to \emph{efficient Nash equilibria} can be achieved, i.e., locally stable outcomes that also maximize a centralized objective. One large class of payoff-based learning dynamics that has been utilized for convergence to Nash equilibria is \emph{reinforcement-based learning}. It may appear under alternative forms, including \emph{discrete-time replicator dynamics} \cite{Arthur93}, \emph{learning automata} \cite{Tsetlin73,Narendra89} or \emph{approximate policy iteration} or \emph{$Q$-learning} \cite{hu_nash_2003}. It is highly attractive to several engineering applications, since agents do not need to know neither the actions of other agents, nor their own utility function. For example, it has been utilized for system identification and pattern recognition \cite{ThathacharSastry04}, distributed network formation and resource-allocation problems \cite{ChasparisShamma11_DGA}.

In reinforcement-based learning, deriving conditions for convergence to Nash equilibria may not be a trivial task especially in the case of large number of agents. Especially in the context of \emph{coordination games} (e.g.,~\cite{ChasparisAriShamma13_SIAM}), two main difficulties are encountered: a) excluding convergence to pure strategies that are \emph{not} Nash equilibria, and b) excluding convergence to mixed strategy profiles. 
Recent work by the author in \emph{perturbed learning automata}  \cite{chasparis_stochastic_2017}, overcame these limitations by directly characterizing the stochastically stable states of the induced Markov chain (independently of the number of players or actions). This type of analysis allowed for acquiring convergence guarantees in multi-player coordination games (thus, extending previous results in reinforcement-based learning restricted only to potential games).

Although Nash equilibria are stochastically stable in coordination games under perturbed learning automata, not all Nash equilibria may be desirable. An example may be drawn from the classical Stag-Hunt coordination game of Table~\ref{Tb:SHG}. 
%
\begin{table}[!ht]
\centering
\caption{The Stag-Hunt Game}
\begin{game}{2}{2}
 & A & B\\
A &$5,5$ &$1,3$\\
B &$3,1$ &$4,4$
\end{game}
\label{Tb:SHG}
\end{table}
In this game, the first player selects the row of the payoff matrix and the second player selects the column. The first element of the selected entry determines the reward of the row player, and the second element determines the reward of the column player. This game has two pure Nash equilibria, which correspond to the symmetric plays $(A,A)$ and $(B,B)$. Ideally, we would prefer that agents eventually learn to play $(A,A)$ which corresponds to the payoff-dominant (or Pareto-efficient) equilibrium. However, existing results in perturbed learning automata \cite{ChasparisShamma11_DGA,ChasparisShammaRantzer15,chasparis_stochastic_2017} demonstrate that $(B,B)$ may prevail asymptotically with positive probability. The reason lies in the cost that an agent experiences when the other agent deviates from a Nash equilibrium, which captures the notion of \emph{risk dominance} (cf.,~\cite{HarsanyiSelten88}). In fact, $(B,B)$ is the \emph{risk-dominant} equilibrium in the Stag-Hunt game of Table~\ref{Tb:SHG}.


In the present paper, we extend the perturbed learning automata dynamics presented in \cite{chasparis_stochastic_2017} to incorporate agents' satisfaction levels, namely \emph{aspiration-based perturbed learning automata} (APLA). In this extended version, an agent reinforces an action based on both repeated selection and its satisfaction level. We provide a stochastically stability analysis of the proposed dynamics in multi-player coordination games. Furthermore, we show that payoff-dominant Nash equilibria are the only stochastically-stable states, as opposed to standard learning automata. 

This paper (also in combination with \cite{chasparis_stochastic_2017}) provides an analytical framework that significantly expands the utility of reinforcement-based learning in strategic-form games. Note though that several classes of \emph{aspiration-based learning} also guarantee convergence to efficient outcomes in coordination games. For example, the \emph{baseline-based} dynamics of \cite{marden_payoff_2009}, the \emph{mode-based} dynamics of \cite{marden_pareto_2014}, the \emph{trial-and-error} dynamics of \cite{young_learning_2009}, and the \emph{aspiration-learning} dynamics of \cite{ChasparisAriShamma13_SIAM} also guarantee convergence to efficient Nash equilibria in certain classes of coordination and weakly-acyclic games. However, existing analysis does not take into account the possibility of noisy observations (with the exception of \cite{marden_payoff_2009} and through the introduction of sufficiently large \emph{exploration phases}). In comparison with these learning dynamics, learning automata can naturally incorporate noisy observations, as demonstrated in the robustness convergence analysis of \cite{HopkinsPosch05}, due to the indirect filtering of measurement noise in the formulation of the agents' strategies.  

In the remainder of the paper, Section~\ref{sec:CoordinationGames} introduces coordination games and Section~\ref{sec:APLA} presents the \emph{aspiration-based perturbed learning automata} dynamics. Section~\ref{sec:StochasticStability} presents the main weak-convergence result and Section~\ref{sec:TechnicalDerivation} its technical derivation. Section~\ref{sec:StochasticallyStableStates} provides a refinement of stochastically stable states together with a simulation study. Finally, Section~\ref{sec:Conclusions} presents concluding remarks.

{\bf Notation:}
\begin{itemize}
\item For a Euclidean topological space, $\magn{\cdot}$ denotes the Euclidean distance.
\item $e_j$ denotes the \emph{unit vector} in $\mathbb{R}^{n}$ where its $j$th entry is equal to 1 and all other entries is equal to 0.
\item $\Delta(n)$ denotes the \emph{probability simplex} of dimension $n$, i.e.,
$\Delta(n) \df \left\{ x\in\mathbb{R}^{n} : x\geq{0}, \mathbf{1}\tr x=1 \right\}.$

\item $\Dirac{x}$ denotes the Dirac measure at $x$.

\item For a finite set $A$, $\magn{A}$ denotes its cardinality.

\item Let $\sigma\in\Delta(\magn{A})$ be a finite probability distribution for some finite set $A$. The random selection of an element of $A$ will be denoted ${\rm rand}_{\sigma}[A]$. If $\sigma=(\nicefrac{1}{\magn{A}},...,\nicefrac{1}{\magn{A}})$, i.e., it corresponds to the uniform distribution, the random selection will be denoted by ${\rm rand}_{\rm unif}[A]$.

\end{itemize}

\section{Coordination Games}	\label{sec:CoordinationGames}

We consider the standard setup of finite strategic-form games. There is a finite set of \emph{agents} or \emph{players}, $\cI=\{1,2,\dotsc,n\}$, and each agent has a finite set of actions, denoted by $\cA_{i}$. The set of action profiles is the Cartesian product $\cA\df \cA_{1}\times\dotsb\times\cA_{n}$; $\alpha_{i}\in\cA_{i}$ denotes an \textit{action} of agent $i$; and
$\alpha=(\alpha_{1},\dotsc,\alpha_{n})\in\cA$ denotes the \textit{action profile} or \textit{joint action} of all agents. The \emph{payoff/utility function} of player $i$ is a mapping $u_{i}:\cA\rightarrow\mathbb{R}$. 
An action profile $\alpha^*\in\cA$ is a \emph{(pure) Nash equilibrium} if, for each $i\in\cI$,
\begin{equation}\label{eq:NashCondition}
u_{i}(\alpha_{i}^*,\alpha_{-i}^*) \geq u_{i}(\alpha_{i}',\alpha_{-i}^*)
\end{equation}
for all $\alpha_{i}'\in\cA_{i}$, where $-i$ denotes the complementary set $\cI\setminus\{i\}$. We denote the set of pure Nash equilibria by $\cA^*$. 

Before defining coordination games, we first need to define the notion of \emph{best response}.
\begin{definition}[Best~Response] \label{Df:BestResponse}
The best response of agent $i\in\cI$ to an action profile $\alpha=(\alpha_{i},\alpha_{-i})\in\cA$ is a set valued map $\mathrm{BR}_{i}:\cA\to 2^{\cA_{i}}$ such that $${\rm BR}_i(\alpha)\df\arg\max_{a\in\cA_i}u_i(a,\alpha_{-i}).$$
\end{definition}

A \emph{coordination game} is defined as follows:

\begin{definition}[Coordination game]\label{def:CoordinationGame}
A game of two or more agents is a coordination game if the following conditions are satisfied:
\begin{enumerate}

\item[\upshape{(a)}]
for any $\alpha\in\cA\setminus\cA^*$, there exist $i\in\cI$ and action $\alpha_{i}'\in \mathrm{BR}_{i}(\alpha)$ such that 
\begin{equation}\label{eq:Condition1CoordinationGame}
u_{j}(\alpha_{i}',\alpha_{-i})\geq u_{j}(\alpha_{i},\alpha_{-i})\,, \text{~for all~} j\neq{i}\,;
\end{equation}

\item[\upshape{(b)}]
for any
$\alpha^*\in\cA^*$, there exist an agent $i$ and an action $\tilde{\alpha}_i\in\cA_i$
such that
\begin{equation}	\label{eq:Condition2CoordinationGame}
u_{j}(\tilde{\alpha}_{i},\alpha^*_{-i})< u_{j}(\alpha^*) \,, \text{~for all~} j\in\cI\,.
\end{equation}

\end{enumerate}
\end{definition}

The conditions of a coordination game establish a weak form of ``coincidence of interests'' among players. For example, condition~(\ref{eq:Condition1CoordinationGame}) states that there always exists a best response of a player that can make no other player worse off. Due to this condition, a pure Nash equilibrium always exists. Furthermore, condition~(\ref{eq:Condition2CoordinationGame}) states that at a pure Nash equilibrium, there exists an action profile that can make every player worse off. It is straightforward to show that coordination games are weakly acyclic games (cf.,~\cite{Young04}). For example, the Stag-Hunt game of Table~\ref{Tb:SHG} satisfy the properties of Definition~\ref{def:CoordinationGame}. Alternative examples can be found, e.g., the \emph{network formation games} and \emph{common-pool games} presented in \cite{ChasparisAriShamma13_SIAM}.

\textit{\textbf{For the remainder of the paper}}, we will be concerned with coordination games that satisfy the \emph{\textbf{Positive-Utility Property}}. 

\begin{property}[Positive-Utility Property]		\label{P:PositiveUtilityProperty}
For any agent $i\in\mathcal{I}$ and any action profile $\alpha\in\mathcal{A}$, $\reward{i}{\alpha}>0$.
\end{property}

\section{Aspiration-based Perturbed Learning Automata (APLA)}	\label{sec:APLA}

In this section, we present a novel reinforcement-based learning algorithm, namely \emph{aspiration-based perturbed learning automata} (APLA). 

The proposed dynamics is presented in Table~\ref{Tb:ReinforcementLearning} and extends the recently developed \emph{perturbed learning automata} \cite{ChasparisShamma11_DGA,chasparis_stochastic_2017}. 
At the first step, each agent $i$ selects an action according to a finite probability distribution (i.e., \emph{strategy}) $x_i(t)\in\cX_i\df\Delta(\magn{\cA_i})$ (capturing its beliefs about the most rewarding action). Its selection is slightly perturbed by a perturbation (or \emph{mutations}) factor $\lambda>0$, such that, with a small probability $\lambda$ agent $i$ follows a uniform strategy (or, it \emph{trembles}).  At the second step, agent $i$ evaluates its new selection by collecting a utility measurement, while in the third step, agent $i$ updates its strategy vector given its new experience. Finally, each agent $i$ updates its discounted running average performance, namely \emph{aspiration level}, $\rho_i\in[\underline{\rho},\overline{\rho}]$, for some $\underline{\rho}$, $\overline{\rho}>0$.
\begin{table}[t!]
\caption{Aspiration-based Perturbed Learning Automata (APLA)}
\boxed{
\begin{minipage}{0.48\textwidth}
At fixed time instances $t=1,2,...$, and for each agent $i\in\cI$, the following steps are executed recursively. Let $\alpha_i(t)$ and $x_i(t)$ denote the current action and strategy of agent $i$, respectively. 
\begin{enumerate}
\item (\emph{\textbf{action update}}) Agent $i\in\cI$ selects a new action $\alpha_i(t+1)$ as follows: 
\begin{eqnarray}	\label{eq:ActionUpdate}
\alpha_i(t+1) = \begin{cases}
{\rm rand}_{x_i(t)}[\mathcal{A}_i], & \mbox{ with probability } 1-\lambda, \cr
{\rm rand}_{\rm unif}[\mathcal{A}_i], & \mbox{ with probability } \lambda,
\end{cases}
\end{eqnarray} 
for some small perturbation factor $\lambda>0$.

\item (\emph{\textbf{evaluation}}) Agent $i$ applies its new action $\alpha_i(t+1)$ and receives a measurement of its utility function $\reward{i}{\alpha(t+1)}>0$. 

\item (\emph{\textbf{strategy update}}) Agent $i$ revises its strategy vector $x_i\in\Delta(\magn{\mathcal{A}_i})$ as follows: 
\begin{eqnarray}	\label{eq:ReinforcementLearningModel}
x_i(t+1) & = & x_i(t) + \epsilon \cdot \reward{i}{\alpha(t+1)} \cdot [e_{\alpha_i(t+1)} - x_i(t)]\cdot \cr &&  \phi(\reward{i}{\alpha(t+1)}-\rho_i(t)) \cr & \df & \mathcal{R}_{i}(\alpha(t+1),x_i(t),\rho_i(t)),
\end{eqnarray}
for some constant step size $\epsilon>0$, 
where
\begin{eqnarray}	\label{eq:SigmoidFunction}
\phi(y) \df 
\begin{cases}
1\,, & \text{if~} y\ge0\,,\\
\max(h,1+y/h)\,, & \text{if~}y<0\,.\end{cases}
\end{eqnarray}
for some positive constant $h>0$. 

\item (\emph{\textbf{aspiration-level update}}) Agent $i$ revises its aspiration level $\rho_i\in[\underline{\rho},\overline{\rho}]$ as follows:
\begin{eqnarray}	\label{eq:AspirationUpdate}
\rho_i(t+1) & = & \rho_i(t) + \nu \cdot \left[ \reward{i}{\alpha(t+1)}-\rho_i(t) \right] \cr
& \df & \mathcal{K}_i(\alpha(t+1),\rho_i(t)).
\end{eqnarray}
for some constant step size $\nu>0$.
\end{enumerate}
\end{minipage}
}
\label{Tb:ReinforcementLearning}
\end{table}

Here we identify actions $\mathcal{A}_i$ with vertices of the simplex, $\{e_1,...,e_{\magn{\mathcal{A}_i}}\}$. For example, if agent $i$ selects its $j$th action at time $t$, then $e_{\alpha_i(t)}\equiv e_j$. To better clarify how the strategies evolve, consider the following toy example. Let the current strategy of player $i$ be $x_i(t) = (\begin{array}{cc} \nicefrac{1}{2} & \nicefrac{1}{2} \end{array})\tr$, i.e., player $i$ has two actions, each assigned probability $\nicefrac{1}{2}$. Let also $\alpha_i(t+1)=1$, i.e., player $i$ selects the first action according to rule (\ref{eq:ActionUpdate}). Then, the new strategy vector for agent $i$, updated according to rule (\ref{eq:ReinforcementLearningModel}), is:
\begin{eqnarray*}
\lefteqn{x_i(t+1) = }\cr && \nicefrac{1}{2} \left(\begin{array}{c} 1 + \epsilon \cdot u_i(\alpha(t+1)) \cdot \phi(u_i(\alpha(t+1))-\rho_i(t))\\ 1 - \epsilon \cdot u_i(\alpha(t+1)) \cdot \phi(u_i(\alpha(t+1))-\rho_i(t)) \end{array}\right).
\end{eqnarray*}
If $y\df u_i(\alpha(t+1))-\rho_i(t) \geq{0}$, i.e., player $i$ receives a satisfactory performance, then $\phi(y)=1$, and the strategy of the selected action is going to increase proportionally to the reward received from this action. If, instead, $y<{0}$, then $\phi(y)=\max\{h,1+y/h\}<{1}$, and the strategy of the selected action is going to increase proportionally to both the reward received and $\phi(y)$. By adjusting $h>0$, we may control the increase in the strategy of a dissatisfactory action, since, if the current reward is far below the current aspiration level, then $\phi(y)=h$.

The introduction of the level of satisfaction captured by $\phi(\cdot)$ in the strategy update is the main contribution of this paper as compared to the original \emph{perturbed learning automata} (PLA) introduced in \cite{ChasparisShamma11_DGA,chasparis_stochastic_2017}, where $h \equiv{1}$. 


Note that by letting the step-size $\epsilon$ to be sufficiently small and since the utility function $\reward{i}{\cdot}$ is uniformly bounded in $\mathcal{A}$, $x_i(t)\in\Delta(\magn{\mathcal{A}_i})$ for all $t$. 

We also deliberately set the step size of the aspiration-level update $\nu>0$ to be different than the step size of the strategy vector update $\epsilon>0$. In general, and for reasons that will become more clear in the forthcoming Section~\ref{sec:TechnicalDerivation}, we would like $\nu$ to be sufficiently larger than $\epsilon$ in order for the aspiration level to evolve at a faster rate than the strategy vector. \emph{\textbf{For the remainder of the paper}}, we will assume the following design property.

\begin{property}	\label{P:FasterAspirationLevel}
Given $\epsilon>0$, we set the step size $\nu=\nu(\epsilon)$ sufficiently small such that, for any $\delta>0$, 
\begin{eqnarray}	\label{eq:StepSizeProperty}
\sup_{i\in\cI}\frac{\log\left(\delta /\magn{\overline{\rho}-\underline{\rho}}\right)}{\log(1-\nu)} \leq \inf_{\alpha\in\cA}\inf_{i\in\cI}\frac{\log(\delta)}{\log(1-\epsilon u_i(\alpha))}
\end{eqnarray}
\end{property}

The ratio of the l.h.s. represents the minimum number of steps that the aspiration-level update of a player $i$ needs to reach a $\delta$-neighborhood of the utility $u_i(\alpha)$ starting from any other action profile. The ratio of the r.h.s. represents the minimum number of steps that the strategy update of a player $i$ needs to reach a $\delta$-neighborhood of the pure strategy vector corresponding to $\alpha$, when playing $\alpha$ continuously. It is clear that for a finite number of actions and bounded utilities, and for $\nu$ sufficiently larger than $\epsilon$, property (\ref{eq:StepSizeProperty}) will be satisfied.

\section{Stochastic Stability} \label{sec:StochasticStability}

\subsection{Terminology and notation}	\label{sec:Terminology}

Let $\cZ\df \mathcal{A}\times \cX \times \mathcal{W}$, where $\cX\df\cX_1\times\ldots\times\cX_n$ and $\mathcal{W}\df[\underline{\rho},\overline{\rho}]^{n}$, i.e., tuples of joint actions $\alpha$, strategy profiles $x=(x_1,...,x_n)$ and aspiration-level profiles $\rho=(\rho_1,...,\rho_n)$. We will denote the elements of the state space $\cZ$ by $z$. 

The set $\mathcal{A}$ is endowed with the discrete topology, $\cX$ and $\mathcal{W}$ with the usual Euclidean topology, and $\cZ$ with the corresponding product topology. We also let $\Bor(\cZ)$ denote the Borel $\sigma$-field of $\cZ$, and $\mathfrak{P}(\cZ)$ the set of \emph{probability measures} (p.m.) on $\Bor(\cZ)$ endowed with the Prohorov topology, i.e., the topology of weak convergence. The dynamics of Table~\ref{Tb:ReinforcementLearning} defines an $\cZ$-valued Markov chain. Let $P_{\lambda}:\cZ\times\Bor(\cZ)\to[0,1]$ denote its transition probability function (t.p.f.), parameterized by $\lambda>0$. We will refer to this process as the \emph{perturbed process}. 

Note that under the perturbed process $P_{\lambda}$ one or more agents may \emph{tremble} (i.e., select randomly an action according to the uniform distribution). Define also the process $P:\cZ\times\Bor(\cZ)\to[0,1]$ where \emph{at most one agent may tremble}. We will refer to this process as the \emph{unperturbed process}.

We let $C_b(\cZ)$ denote the Banach space of real-valued continuous functions on $\cZ$ under the sup-norm (denoted by $\|\cdot\|_{\infty}$) topology. For $f\in\Cc(\cZ)$, define
\begin{equation*}
P_{\lambda}f(z) \df \int_{\cZ}P_{\lambda}(z,dy)f(y),
\end{equation*}
and 
\begin{equation*}
\mu[f] \df \int_{\cZ}\mu(dz)f(z), \mbox{ for } \mu\in\mathfrak{P}(\cZ).
\end{equation*}

The process governed by the unperturbed process $P$ will be denoted by $Z=\{Z_{t} : t\ge0\}$. Let $\Omega\df\cZ^{\infty}$ denote the canonical path space, i.e., an element $\omega\in\Omega$ is a sequence $\{\omega(0),\omega(1),\dotsc\}$, with $\omega(t)= (\alpha(t),x(t),\rho(t))\in\cZ$. We use the same notation for the elements $(\alpha,x,\rho)$ of the space $\cZ$ and for the coordinates of the process $Z_{t}=(\alpha(t),x(t),\rho(t))$.
Let also $\Prob_{z}[\cdot]$ denote the unique p.m. induced by the unperturbed process $P$ on the product $\sigma$-field of $\cZ^{\infty}$, initialized at $z=(\alpha,x,\rho)$. 

\subsection{Stochastic stability}

First, note that both $P$ and $P_{\lambda}$ ($\lambda,h>0$) satisfy the \emph{weak-Feller property} (cf.,~\cite[Definition~4.4.2]{Lerma03}).
\begin{proposition}		\label{Pr:WeakFeller}
Both the unperturbed process $P$ ($\lambda=0$) and the perturbed process $P_{\lambda}$ ($\lambda>0$) satisfy the weak-Feller property.
\end{proposition}
\ifproofs
\begin{proof}
Let us consider the perturbed process $P_{\lambda}$. The proof for the unperturbed process will be directly implied by employing $\lambda=0$. Let us also consider any sequence $\{z^{(k)}=(\alpha^{(k)},x^{(k)},\rho^{(k)})\}$ such that $z^{(k)}\to{z}=(\alpha,x,\rho)\in\cZ$. 

For any open set $O\in\Bor(\cZ)$, the following holds:
\begin{eqnarray*}
\lefteqn{P_{\lambda}(z^{(k)}=(\alpha^{(k)},x^{(k)},\rho^{(k)}),O) }\cr 
& = & \sum_{\alpha\in\mathcal{P}_{\cA}(O)}\Big\{\prod_{i\in\cI}\tilde{x}_{i\alpha_i}^{(k)} \cdot \cr 
&& \prod_{i\in\cI}\Prob_{z^{(k)}}\left[\mathcal{R}_{i}(\alpha,x_i^{(k)},\rho_i^{(k)})\in\mathcal{P}_{\cX_i}(O)\right] \cdot \cr &&
\prod_{i\in\cI}\Prob_{z^{(k)}}\left[\mathcal{K}_{i}(\alpha,\rho_i^{(k)})\in\mathcal{P}_{\mathcal{W}_i}(O)\right]\Big\} \cr
& = & \sum_{\alpha\in\mathcal{P}_{\mathcal{A}}(O)} \Big\{\prod_{i=1}^{n} \Big\{ \mathbb{I}_{\mathcal{P}_{\cX_i}(O)}\left(\mathcal{R}_i(\alpha,x_i^{(k)},\rho_i^{(k)})\right) \cdot\cr && \mathbb{I}_{\mathcal{P}_{\mathcal{W}_i}(O)}\left(\mathcal{K}_i(\alpha,\rho_i^{(k)})\right) \cdot \tilde{x}_{i\alpha_i}^{(k)}\Big\}\Big\},
\end{eqnarray*}
where $\mathcal{P}_{\cX_i}(O)$, $\mathcal{P}_{\mathcal{A}}(O)$ and $\mathcal{P}_{\mathcal{W}_i}(O)$ are the \emph{canonical projections} defined by the product topology, and 
$$\tilde{x}_{i\alpha_i}^{(k)}\df (1-\lambda)x_{i\alpha_i}^{(k)} + {\lambda}/{\magn{\cA_i}}.$$ Similarly, we have: 
\begin{eqnarray*}
\lefteqn{P_{\lambda}(z,O)} \cr & = & \sum_{\alpha\in\mathcal{P}_{\mathcal{A}}(O)} \Big\{\prod_{i=1}^{n}\Big\{\mathbb{I}_{\mathcal{P}_{\cX_i}(O)}\left(\mathcal{R}_i\left(\alpha,x_i,\rho_i\right)\right) \cdot \cr && \mathbb{I}_{\mathcal{P}_{\cX_i}(O)}\left(\mathcal{R}_i\left(\alpha,x_i,\rho_i\right)\right) \cdot \tilde{x}_{i\alpha_i}\Big\}.
\end{eqnarray*}

To investigate the limit of $P_{\lambda}(Z^{(k)},O)$ as $k\to\infty$, it suffices to investigate the behavior of the sequences $$\beta_i^{(k)} \df \mathbb{I}_{\mathcal{P}_{\cX_i}(O)}\left(\mathcal{R}_i(\alpha,x_i^{(k)},\rho_i^{(k)}\right),$$ and $$\gamma_i^{(k)} \df \mathbb{I}_{\mathcal{P}_{\mathcal{W}_i}(O)}\left(\mathcal{K}_i(\alpha,\rho_i^{(k)}\right).$$ 

Let us first investigate the sequence $\beta_i^{(k)}$. We distinguish the following (complementary) cases:

(a) $\mathcal{R}_i(\alpha,x_i,\rho_i)\notin\mathcal{P}_{\cX_i}(O)$ and $\mathcal{R}_i(\alpha,x_i,\rho_i)\notin\partial\mathcal{P}_{\cX_i}(O)$: In this case, there exists an open ball about the next strategy vector that does not share any common points with the canonical projection of $O$. Due to the continuity of the function $\mathcal{R}_i(\alpha,\cdot, \cdot)$, we have that $\beta_i^{(k)}\to \beta_i \df \mathbb{I}_{\mathcal{P}_{\cX_i}(O)}(\mathcal{R}_i(\alpha,x_i,\rho_i))\equiv{0}$.

(b) $\mathcal{R}_i(\alpha,x_i,\rho_i)\in\mathcal{P}_{\cX_i}(O)$: In this case, there exists an open ball about the next strategy vector that belongs to the canonical projection of $O$, since $O\in\Bor(\cZ)$. Due to the continuity of the function $\mathcal{R}_i(\alpha,\cdot,\cdot)$ with respect to both the strategy $x_i$ and the aspiration level $\rho_i$, we have that $\beta_i^{(k)}\to \beta_i=1$.

(c) $\mathcal{R}_i(\alpha,x_i,\rho_i)\notin\mathcal{P}_{\cX_i}(O)$ and $\mathcal{R}_i(\alpha,x_i,\rho_i)\in\partial\mathcal{P}_{\cX_i}(O)$: In this case,  $\beta_i \equiv{0}$. We conclude that $\liminf_{k\to\infty}{\beta_i^{(k)}}\geq \beta_i = 0$, since $\beta_i^{(k)}\in\{0,1\}$.

In either one of the above (complementary) cases (a), (b) or (c), we have that $\liminf_{k\to\infty}{\beta_i^{(k)}}\geq \beta_i$. Following the exact same reasoning, and the continuity of the mapping $\mathcal{K}_i(\alpha,\cdot)$ with respect to the aspiration level $\rho_i$, we also derive that $\liminf_{k\to\infty}\gamma_i^{(k)}\geq\gamma_i$ (for all the corresponding (a), (b) and (c) cases).

Finally, due to the continuity of the perturbed strategy vector $\tilde{x}_{i\alpha_i}$ with respect to $x_{i\alpha_i}$, we conclude that for any sequence $z^{(k)}\to z$, $$\liminf_{k\to\infty}P_{\lambda}(z^{(k)},O) \geq P_{\lambda}(z,O).$$ By \cite[Proposition~7.2.1]{Lerma03}, we conclude that $P_{\lambda}$ satisfies the weak-Feller property.

The above derivation can be generalized to any selection probability function $f(x_{i\alpha_i})$ in the place of $\tilde{x}_{i\alpha_i}$, provided that it is a continuous function. Thus, the proof for the unperturbed process $P$ follows the exact same reasoning by simply setting $f(x_{i\alpha_i})=x_{i\alpha_i}$. 
\end{proof}
\fi

The measure $\mu_{\lambda}\in\mathfrak{P}(\cZ)$ is called an \emph{invariant probability measure} (i.p.m.) for $P_{\lambda}$ if
\begin{equation*}
(\mu_{\lambda}P_{\lambda})(A) \df \int_{\cZ}\mu_{\lambda}(dz)P_{\lambda}(z,A) = \mu_{\lambda}(A), \qquad A\in\Bor(\cZ).
\end{equation*}
Since $\cZ$ defines a locally compact separable metric space and $P$, $P_{\lambda}$ satisfy the weak-Feller property, they both admit an i.p.m., denoted $\mu$ and $\mu_{\lambda}$, respectively \cite[Theorem~7.2.3]{Lerma03}.

We would like to characterize the \emph{stochastically stable states} $z\in\cZ$ of $P_{\lambda}$, that is any state $z\in\cZ$ for which any collection of i.p.m. $\{\mu_{\lambda}\in\mathfrak{P}(\cZ):\mu_{\lambda}P_{\lambda}=\mu_{\lambda},\lambda>0\}$ satisfies $\liminf_{\lambda\to{0}}\mu_{\lambda}(z)>0$. As the forthcoming analysis will show, the stochastically stable states will be a subset of the set of \emph{pure strategy states} (p.s.s.) defined as follows:
\begin{definition}[Pure Strategy State]	\label{def:PureStrategyState}
A pure strategy state is a state $s=(\alpha,x,\rho)\in\cZ$ such that for all $i\in\mathcal{I}$, $x_i = e_{\alpha_i}$ and $\rho_i=u_i(\alpha)$, i.e., $x_i$ coincides with the vertex of the probability simplex $\Delta(\magn{\mathcal{A}_i})$ which assigns probability 1 to action $\alpha_i$, and $\rho_i$ coincides with the utility of agent $i$ under action profile $\alpha$.
\end{definition}

We will denote the set of pure strategy states by $\mathcal{S}$. Pure strategy states that correspond to pure Nash equilibria, will be referred to as \emph{pure Nash equilibrium states} and will be denoted by $\cS^*$. For any pure strategy state $s^*=(\alpha^*,x^*,\rho^*)$, define the $\delta$-neighborhood of $s^*$ as follows
\begin{align*}
& \mathcal{N}_{\delta}(s^*)\df \cr
& \{z=(\alpha,x,\rho)\in\cZ:\alpha=\alpha^*\,, |x-x^*|<\delta\,, |\rho-\rho^*|<\delta\}.
\end{align*}

\begin{theorem}[Stochastic Stability]		\label{Th:StochasticStability}
In any coordination game (Definition~\ref{def:CoordinationGame}) and under the aspiration-based perturbed learning automata of Table~\ref{Tb:ReinforcementLearning} (APLA), there exists a unique probability vector $\pi=(\pi_1,...,\pi_{\magn{\cS^*}})$ such that, for any collection of i.p.m.'s $\{\mu_{\lambda}\in\mathfrak{P}(\cZ):\mu_{\lambda}P_{\lambda}=\mu_{\lambda}, \lambda>0\}$, 
\begin{itemize}
\item[(a)] $\lim_{\lambda\downarrow{0}}\mu_{\lambda}(\cdot) = \hat{\mu}(\cdot) \df \sum_{s\in\cS^*}\pi_s\Dirac{s}(\cdot),$ where convergence is in the weak sense.
\item[(b)] The probability vector $\pi$ is an invariant distribution of the (finite-state) Markov process $\hat{P}$, such that, for any $s,s'\in\cS^*$, 
\begin{equation}	\label{eq:FiniteStateMarkovChain}
\hat{P}_{ss'} \df \lim_{t\to\infty} QP^t(s,\mathcal{N}_{\delta}(s')),
\end{equation}
for any $\delta>0$ sufficiently small, where $Q$ is the t.p.f. corresponding to \emph{at least two players trembling} (i.e., following the uniform distribution of (\ref{eq:ActionUpdate})).
\end{itemize}
\end{theorem}

Theorem~\ref{Th:StochasticStability} establishes weak convergence of the i.p.m. of $P_\lambda$ with the invariant distribution of a finite Markov chain $\hat{P}$, whose support is on the set of pure Nash equilibrium states. Thus, from the ergodicity of $\mu_{\lambda}$, we have that the expected percentage of time that the process spends in any $O\in\Bor(\cZ)$ such that $\partial{O}\cap\cS^*\neq\varnothing$ is given by $\hat{\mu}(O)$ as $h,\lambda\downarrow{0}$ and time increases, i.e.,
\begin{equation*}
\lim_{h,\lambda\downarrow{0}}\left(\lim_{t\to\infty}\;
\frac{1}{t}\sum_{k=0}^{t-1}P_{\lambda}^{k}(x,O)\right) = \Hat{\mu}(O)\,.
\end{equation*}
The methodology for assessing which Nash equilibria are stochastically stable will follow in the forthcoming Section~\ref{sec:StochasticallyStableStates}.


\section{Technical Derivation}	\label{sec:TechnicalDerivation}

In this section, we provide the main steps for the proof of Theorem~\ref{Th:StochasticStability}. We begin by investigating the asymptotic behavior of the unperturbed process $P$, and then we characterize the i.p.m. of the perturbed process with respect to the pure Nash equilibrium states $\cS^*$.

\subsection{Unperturbed Process}	\label{Sc:UnperturbedProcess}

Recall that the unperturbed process with t.p.f. $P$ has been defined such that \emph{at most one agent may tremble}. We first present two technical lemmas that will help us identify the behavior of the unperturbed process. 

Let $\uptau(D)$ denote the first hitting time of the unperturbed process to a set $D\subset\Bor(\cZ)$. 

For some action profile $\alpha\in\cA$ and $\delta>0$, define the set: 
\begin{equation*}
D_{\delta}(\alpha)\df\{z\in\cZ:\rho_i \leq u_i(\alpha) + \delta \,, x_{i\alpha_i}>0\,, \forall{i\in\cI}\}.
\end{equation*}
The set $D_{\delta}(\alpha)\subset\Bor(\cZ)$, corresponds to any state $z\in\cZ$ at which the aspiration level is below $u_i(\alpha)+\delta$ for some given $\alpha\in\cA$ and $\delta>0$.
Define also the event $$\Gamma_{\delta,t}\df\{\exists\alpha\in\cA : \uptau(D_{\delta}(\alpha)) < t \},$$ i.e., the event $\Gamma_{t}$ corresponds to the case that the process first reaches $D_{\delta}(\alpha)$ for some action profile $\alpha$ before time $t$. 

The first lemma states that, for any initial state $z$, at least one $\Gamma_{\delta,t}$ occurs for some $t>0$.
\begin{lemma}	\label{Lm:ConvergenceToDdelta}
For any $\delta>0$, and any $z=(\alpha,x,\rho)\in\cZ$, $\Prob_z[\Gamma_{\delta,\infty}] = 1.$
\end{lemma}
\ifproofs
\begin{proof}
Consider any initial state $z=(\cdot,x(0),\rho(0))$. Since $x_i\in\cX_i\equiv\Simplex{\magn{\cA_i}}$, there exists an action profile $\alpha\in\cA$ such that $x_{i\alpha_i}(0)>0$ for all $i\in\cI$. For some $\delta>0$, define $T(\delta)$ to be the maximum (with respect to $\cA$) number of iterations required for the aspiration level profile $\rho$ to drop from $\overline{\rho}$ (i.e., its maximum value) to $D_{\delta}(\alpha)$, when playing only action profile $\alpha$. Let also $D_{\delta}(\alpha)^{c}$ denote the complement of $D_{\delta}(\alpha)$. We will first consider the case that the initial state $z\in D_{\delta}(\alpha)^c$. In this case, we have:
\begin{eqnarray*}
\lefteqn{\Prob_{z}[\uptau(D_{\delta}(\alpha))<\infty]} \cr 
& \geq & \Prob_{z}[\alpha(t)=\alpha\,, \forall t\leq T(\delta)|\rho_i(0)=\overline{\rho}\,,\forall{i}] \cr
& \geq & \prod_{k=1}^{T(\delta)} (1-\lambda) \prod_{i=1}^{n} x_{i\alpha_i}(0) \cr
& \df & \kappa(\lambda).
\end{eqnarray*}
The first inequality results from the fact that one possible sample path that reaches $D_{\delta}(\alpha)$ corresponds to playing action $\alpha$ continuously, and the probability of this path is smaller when we start from $\overline{\rho}$. The second inequality results from the fact that, under the unperturbed process $P$, only one agent may tremble at any given time. Finally, note that by selecting action $\alpha$ continuously for $T(\delta)$ steps, $x_{i\alpha_i}(t)\geq x_{i\alpha_i}(0)$ for all $0\leq{t}\leq T(\delta)$. By selecting $\nu>0$ (according to Property~\ref{P:FasterAspirationLevel}), $T(\delta)$ is finite, and $\kappa(\lambda)>0$. Finally, note that if, instead, $z\in D_{\delta}(\alpha)$, then $\Prob_{z}[\uptau(D_{\delta}(\alpha))<\infty]=1$. Thus, we conclude that $$\inf_{z}\Prob_{z}[\uptau(D_{\delta}(\alpha))<\infty]\geq\kappa(\lambda)>0.$$

For some $t > T(\delta)$, let us define the event:
\begin{eqnarray*}
\lefteqn{ C_{\delta,t} \df \{ \exists{k}\leq{t}, \alpha'\in\cA: \alpha(\tau)=\alpha'\,, } \cr && \forall\tau : k < \tau \leq k + T(\delta) \leq t\}.
\end{eqnarray*}
In other words, $C_{\delta,t}$ corresponds to the event that some action profile $\alpha'$ has been selected (continuously) for at least $T(\delta)$ times before time $t$. Note that $C_{\delta,t}\subseteq C_{\delta,t+1}$. Furthermore,
$\inf_{z}\Prob_{z}[C_{\delta,t+T(\delta)}|C_{\delta,t}^{c}] = \kappa(\lambda) > 0$. Thus, from the counterpart of the Borel-Cantelli Lemma (cf.,~\cite[Lemma~1]{bruss_counterpart_1980}), and continuity from below, we conclude that $\inf_{z}\Prob_z[C_{\delta,\infty}]=1$, i.e., the probability that at least one $C_{\delta,t}$ occurs, starting from any $z$, is one. Given that $C_{\delta,t}\subseteq \Gamma_{\delta,t}$ for any $t$, the conclusion follows.
\end{proof}
\fi

The second lemma states that the unperturbed process reaches (infinitely often) a state at which the aspiration level is below the utility level of a Nash equilibrium and its strategy assigns positive probability to it. Define the event:
$$\Gamma_{t}^*\df\{\exists\alpha^*\in\cA^*\,:\uptau(D_{0}(\alpha^*))<t\},$$ which corresponds to the case that the aspiration level is below the utility level of a Nash equilibrium.
\begin{lemma}	\label{Lm:ConvergenceToD}
For any $z=(\alpha,x,\rho)\in\cZ$, $\Prob_{z}[\Gamma_{\infty}^*]=1.$
\end{lemma}
\ifproofs
\begin{proof}
By Lemma~\ref{Lm:ConvergenceToDdelta}, there exists a subsequence $\{t_k\}$ such that $Z(t_k)\in D_{\delta}(\alpha^{(k)})$ for some action profile $\alpha^{(k)}\in\cA$. We distinguish the following (complementary) cases.

(a) $\alpha^{(k)}\in\cA^*$. In this case, $\alpha^{(k)}$ corresponds to a pure Nash equilibrium. By definition of a coordination game, there exists agent $i^*=i^*(\alpha^{(k)})$ and $\tilde{\alpha}_{i^*}$ that makes every other agent worse off. By selecting $\delta=\delta(\nu)>0$ sufficiently small, such drop in the performance brings the aspiration level of every agent strictly below $u_i(\alpha^{(k)})$. Formally, select $\delta<\delta^*$, where
\begin{equation*}
\delta^* \df \frac{\nu}{\nu+1} \inf_{\alpha^*\in\cA^*} \inf_{j\in\cI} |u_j(\tilde{\alpha}_{i^*(\alpha^*)},\alpha_{-i}^*)-u_j(\alpha^*))| > 0.
\end{equation*}
Then,
\begin{align*}
& \Prob_{z}[Z(t_k+1)\in D_0(\alpha^{*})|Z(t_k)\in D_{\delta}(\alpha^{(k)})] \cr & \geq \frac{\lambda}{\magn{\cA_{i^*(\alpha^{(k)})}}} \df \gamma_{1}(\lambda),
\end{align*}
which is strictly positive since $\lambda>0$.

(b) $\alpha^{(k)}\in\cA\backslash\cA^*$. In this case, $\alpha^{(k)}$ corresponds to an action profile that is not a Nash equilibrium. According to the definition of a coordination game, there exists a finite sequence of action profiles starting from $\alpha$, namely $\alpha^{(0)},\alpha^{(1)},...,\alpha^{(L)}$, such that: a) $\alpha^{(0)}=\alpha$, b) $\alpha^{(L)}\in\cA^*$ and c) for each $\ell=1,...,L$, there exists $i$ such that $\alpha_i^{(\ell+1)}\in {\rm BR}_i(\alpha^{(\ell)})$ satisfies condition (\ref{eq:Condition1CoordinationGame}). Thus, there exists finite integer $L>0$, such that 
\begin{align*}
& \Prob_{z}[Z(t_k+L)\in D_0(\alpha^{*})|Z(t_k)\in D_{\delta}(\alpha^{(k)})] \cr & \geq \prod_{\ell=1}^{L}\frac{\lambda}{\magn{\cA_{\ell}}}\prod_{i\in\cI\backslash{\ell}}x_{i\alpha^{(\ell)}}(t_k+\ell) \df \gamma_{2}(\lambda,\epsilon).
\end{align*}
Along this sequence of best responses there is no agent that gets worse off. Thus, along this sample path, $x_{i\alpha^{(\ell)}}$ increases with an order of $\epsilon$ (and independent of $h$) for every agent $i$. Thus, $\gamma_2(\lambda,\epsilon)>0$.

We can define a subsequence $\{t_{k(m)}\}_m$, that takes into account the size of $L$, such that $\Gamma_{t_{k(m)}}\subseteq \Gamma_{t_{k(m+1)}}$ and $$\inf_{z}\Prob_{z}[\Gamma_{t_{k(m+1)}}|\Gamma_{t_{k(m)}}^{c}]\geq\min\{\gamma_1,\gamma_2\}>0.$$ Thus, the conclusion follows from the counterpart of the Borel-Cantelli Lemma (cf.,~\cite[Lemma~1]{bruss_counterpart_1980}).
\end{proof}
\fi

Next, we will use Lemma~\ref{Lm:ConvergenceToD} to show that the process will reach a $\delta$-neighborhood of a pure Nash equilibrium infinitely often with probability one. 

For any $\delta>0$, define the event:
\begin{equation*}
B_{\delta,t} \df \left\{\exists s^*\in\cS^* \,: \uptau(\mathcal{N}_{\delta}(s^*)) < t \right\}.
\end{equation*}
In other words, $B_{\delta,t}$ corresponds to the event that the unperturbed process has reached a $\delta$-neighborhood of a pure Nash equilibrium state before time instance $t$. \begin{proposition}	\label{Pr:ConvergenceToPureNashEquilibria}
For any $\delta>0$ and any initial state $z=(\alpha,x,\rho)\in\cZ$, 
$\Prob_{z}[B_{\delta,\infty}]= 1.$
\end{proposition}
\ifproofs
\begin{proof}
By Lemma~\ref{Lm:ConvergenceToD}, there exists a subsequence $\{t_k\}$ such that $Z(t_k)\in D_{0}(\alpha^{(k)})$ for some pure Nash equilibrium $\alpha^{(k)}\in\cA^*$. Let us consider one such $t_k$, for some $k$. Let us also consider a sample path of the unperturbed process, where action $\alpha^{(k)}$ is played continuously until $\mathcal{N}_{\delta}(s^{(k)})$ is reached, where $s^{(k)}$ is the pure strategy state corresponding to action profile $\alpha^{(k)}$. Let $T^*(\delta)$ be the maximum (with respect to $\cA$) number of steps required for the process to reach $\mathcal{N}_{\delta}(s^{(k)})$ when playing action profile $\alpha^{(k)}$ continuously. Proposition~4.1 in \cite{chasparis_stochastic_2017} shows that such sample path occurs with strictly positive probability (of order of $\epsilon$), say $\gamma(\epsilon)$. 
Then, $$\Prob_{z}[B_{\delta,t_k+T^*(\delta)}|Z(t_k)\in D_0(\alpha^{(k)})] \geq \gamma(\epsilon) > 0.$$ We can define a subsequence $\{t_{k(m)}\}$ such that $t_{k(m+1)}\geq t_{k(m)} + T^*(\delta)$, such that $B_{\delta,t_{k(m)}}\subseteq B_{\delta,t_{k(m+1)}}$ and $$\inf_{z}\Prob_z[B_{\delta,t_{k(m+1)}}|B_{\delta,t_{k(m)}}^c] \geq \gamma(\epsilon) > 0.$$
Thus, from the counterpart of the Borel-Cantelli Lemma (cf.,~\cite[Lemma~1]{bruss_counterpart_1980}), we conclude that $\Prob_{z}[B_{\delta,\infty}]=1.$ 
%
\end{proof}
\fi

Proposition~\ref{Pr:ConvergenceToPureNashEquilibria} states that the unperturbed process will reach a $\delta$-neighborhood of a Nash equilibrium infinitely often with probability one. Note that this derivation is independent of the size of $h$ and $\lambda$.

\begin{proposition}[Limiting t.p.f. of unperturbed process]	\label{Pr:LimitingUnperturbedTPF}
Let $\mu$ denote an i.p.m. of $P$. Then, there exists a t.p.f. $\Pi$ on $\cZ\times\Bor(\cZ)$ with the following properties:
\begin{itemize}
\item[(a)] for $\mu$-a.e. $z\in\cZ$, $\Pi(z,\cdot)$ is an i.p.m. for $P$;
\item[(b)] for all $f\in\Cc(\cZ)$, $\lim_{t\to\infty}\|P^tf-\Pi f\|_{\infty}=0$;
\item[(c)] $\mu$ is an i.p.m. for $\Pi$;
\item[(d)] the support\footnote{The \emph{support} of a measure $\mu$ on $\cZ$ is the unique closed set $F\subset\Bor(\cZ)$ such that $\mu(\cZ\backslash{F})=0$ and $\mu(F\cap{O})>0$ for every open set $O\subset\cZ$ such that $F\cap{O}\neq\varnothing$.} of $\Pi$ is on $\cS^*$ for all $z\in\cZ$.
\end{itemize}
\end{proposition}
\ifproofs
\begin{proof}
The state space $\cZ$ is a locally compact separable metric space and the t.p.f. of the unperturbed process $P$ admits an i.p.m. due to the weak-Feller property. Thus, statements (a), (b) and (c) follow directly from \cite[Theorem~5.2.2 (a), (b), (e)]{Lerma03}. 

(d) Let us assume that the support of $\Pi$ includes points in $\cZ$ other than the pure Nash equilibrium states in $\cS^*$. Then, there exists an open set $O\in\Bor(\cZ)$ such that $O\cap\cS^*=\varnothing$ and $\Pi(z^*,O)>0$ for some $z^*\in\cZ$. According to (b), $P^{t}$ converges weakly to $\Pi$. Thus, from the Portmanteau theorem (cf.,~\cite[Theorem~1.4.16]{Lerma03}), we have that $\liminf_{t\to\infty} P^{t}(z^*,O) \geq \Pi(z^*,O)>0.$ However, this contradicts Proposition~\ref{Pr:ConvergenceToPureNashEquilibria}. 
\end{proof}
\fi

Proposition~\ref{Pr:LimitingUnperturbedTPF} states that the limiting unperturbed t.p.f. converges weakly to a t.p.f. $\Pi$ which accepts the same invariant probability measure as $P$. Furthermore, \emph{the support of $\Pi$ is the set of pure Nash equilibrium states in $\cS^*$}. This is a rather handy property, since the limiting perturbed process can also be ``related'' (in a weak-convergence sense) to the t.p.f. $\Pi$, as it will be shown in the following section.

\subsection{Invariant probability measure (i.p.m.) of perturbed process}

Note that the t.p.f. of the perturbed process can be decomposed as follows:
\begin{equation}	\label{eq:Decomposition}
P_{\lambda} = (1-\varphi(\lambda))P + \varphi(\lambda)Q
\end{equation}
where $Q$ is the t.p.f. of the one-step process where \emph{at least two agents tremble simultaneously}, i.e., they play an action uniformly at random. Note that $$1-\varphi(\lambda) = (1-\lambda)^{n}+n\lambda(1-\lambda)^{n-1}$$ is the probability that \emph{at most one agent trembles}. It is straightforward to check that $\varphi(\lambda)\to{0}$ as $\lambda\downarrow{0}$.



Define also the infinite-step t.p.f. when trembling only at the first step (briefly, \emph{lifted} t.p.f.) as follows: 
\begin{equation}
P_{\lambda}^{L} \df \varphi(\lambda)\sum_{t=0}^{\infty}(1-\varphi(\lambda))^{t}QP^{t} = Q R_{\lambda}
\end{equation}
where
$R_{\lambda} \df \varphi(\lambda)\sum_{t=0}^{\infty}(1-\varphi(\lambda))^{t}P^{t},$
i.e., $R_{\lambda}$ corresponds to the \emph{resolvent} t.p.f. 

In the following proposition, we establish weak-convergence of the lifted t.p.f. $P_{\lambda}^{L}$ with $Q\Pi$ as $\lambda\downarrow{0}$, which will further allow for an explicit characterization of the weak limit points of the i.p.m. of $P_{\lambda}$.

\begin{proposition}[i.p.m. of perturbed process]		\label{Pr:WeakLimitPointsOfPerturbedInvariantMeasures}
The following hold:
\begin{itemize}
\item[(a)] For $f\in\Cc(\cZ)$, $\lim_{\lambda\to{0}}\|R_{\lambda}f-\Pi{f}\|_{\infty} = 0.$
\item[(b)] For $f\in\Cc(\cZ)$, $\lim_{\lambda\to{0}}\|P_{\lambda}^{L}f-Q\Pi{f}\|_{\infty} = 0$.
\item[(c)] Any invariant distribution $\mu_{\lambda}$ of $P_\lambda$ is also an invariant distribution of $P_{\lambda}^{L}$.
\item[(d)] Any weak limit point in $\mathfrak{P}(\cZ)$ of $\mu_{\lambda}$, as $\lambda\to{0}$, is an i.p.m. of $Q\Pi$.
\end{itemize}
\end{proposition}
\ifproofs
\begin{proof}
The proof follows the exact same reasoning with the proof of \cite[Proposition~4.3]{chasparis_stochastic_2017}.
\end{proof}
\fi

Proposition~\ref{Pr:WeakLimitPointsOfPerturbedInvariantMeasures} establishes convergence (in a weak sense) of the i.p.m. $\mu_{\lambda}$ of the perturbed process to an i.p.m. of $Q\Pi$. In the following section, this convergence result will allow for a more explicit characterization of $\mu_{\lambda}$ as $\lambda\downarrow{0}$.

\subsection{Equivalent finite-state Markov process}	\label{sec:FiniteStateMarkovChainEquivalence}

Define the finite-state Markov process $\hat{P}$ as in (\ref{eq:FiniteStateMarkovChain}). 

\begin{proposition} [Unique i.p.m. of $Q\Pi$]		\label{Pr:UniqueInvariantPMofQPi}
There exists a unique i.p.m. $\hat{\mu}$ of $Q\Pi$. It satisfies 
\begin{equation}	\label{eq:InvariantMeasureQP_derivation}
\hat{\mu}(\cdot) = \sum_{s\in\mathcal{S}^*}\pi_s\Dirac{s}(\cdot)
\end{equation}
for some constants $\pi_s\geq{0}$, $s\in\mathcal{S}^*$. Moreover, $\pi=(\pi_1,...,\pi_{\magn{\mathcal{S}}})$ is an invariant distribution of $\hat{P}$, i.e., $\pi=\pi\hat{P}$.
\end{proposition}
\ifproofs
\begin{proof}
From Proposition~\ref{Pr:LimitingUnperturbedTPF}(d), we know that, the support of $\Pi$ is the set of pure Nash equilibrium states $\mathcal{S}^*$. Thus, the support of $Q\Pi$ is also on $\mathcal{S}^*$. From Proposition~\ref{Pr:WeakLimitPointsOfPerturbedInvariantMeasures}, we know that $Q\Pi$ admits an i.p.m., say $\hat{\mu}$, whose support is also $\mathcal{S}^*$. Thus, $\hat{\mu}$ admits the form of (\ref{eq:InvariantMeasureQP_derivation}), for some constants $\pi_{s}\geq{0}$, $s\in\mathcal{S}^*$.

For any two distinct $s,s'\in\cS^*$, note that $\mathcal{N}_{\delta}(s')$, $\delta>0$, is a continuity set of $Q\Pi(s,\cdot)$, i.e., $Q\Pi(s,\partial\mathcal{N}_{\delta}(s'))=0$. Thus, from Portmanteau theorem, given that $QP^{t}\Rightarrow Q\Pi$, $$Q\Pi(s,\mathcal{N}_{\delta}(s')) = \lim_{t\to\infty}QP^{t}(s,\mathcal{N}_{\delta}(s')) = \hat{P}_{ss'}.$$  If we also define $\pi_s \df \hat{\mu}(\mathcal{N}_{\delta}(s))$, then
$$\pi_{s'} = \hat{\mu}(\mathcal{N}_{\delta}(s')) = \sum_{s\in\mathcal{S}}\pi_s Q\Pi(s,\mathcal{N}_{\delta}(s')) = \sum_{s\in\mathcal{S}}\pi_s\hat{P}_{ss'},$$ which shows that $\pi$ is an invariant distribution of $\hat{P}$, i.e., $\pi=\pi\hat{P}$.

It remains to establish uniqueness of the invariant distribution of $Q\Pi$. Note that the set $\mathcal{S}^*$ of pure Nash equilibrium states states is isomorphic with the set $\mathcal{A}^*$ of pure Nash equilibrium profiles. If two or more agents tremble simultaneously (as t.p.f. $Q$ dictates), then all other pure Nash equilibrium profiles in $\mathcal{A}^*$ have positive probability of being reached (due to the fact that $h>0$), i.e., $Q\Pi(\alpha,\alpha')>0$ for all $\alpha'\in\mathcal{A}^*$ and $i\in\cI$. It follows by Proposition~\ref{Pr:ConvergenceToPureNashEquilibria} that $Q\Pi(\alpha,\alpha')>0$ for all $\alpha'\in\cA^*$ and $i\in\cI$. Finite induction then shows that $(Q\Pi)^{n}(\alpha,\alpha')>0$ for all $\alpha,\alpha'\in\cA^*$. It follows that if we restrict the domain of $Q\Pi$ to $\mathcal{S}^*$, it defines an irreducible stochastic matrix. Therefore, $Q\Pi$ has a unique i.p.m.
\end{proof}
\fi

\subsection{Proof of Theorem~\ref{Th:StochasticStability}}	\label{sec:Proof:StochasticStability}

Theorem~\ref{Th:StochasticStability}(a)--(b) is a direct implication of Propositions~\ref{Pr:WeakLimitPointsOfPerturbedInvariantMeasures}--\ref{Pr:UniqueInvariantPMofQPi}.

\section{Convergence to Efficient Outcomes}	\label{sec:StochasticallyStableStates}

\subsection{Payoff-dominant states}	\label{sec:PayoffDominantStates}

According to Theorem~\ref{Th:StochasticStability}, the stochastically stable states is a subset of the set of pure Nash equilibrium states. We may further refine the set of stochastically stable states if we introduce additional structural constraints. One such constraint is the existence of a payoff-dominant set of action profiles.

\begin{definition}[Payoff-dominant states]
The set of payoff-dominant states, denoted by $\overline{\cS}\subseteq\cS^*$, is such that, for any $s^*=(\alpha^*,x^*,\rho^*)\in\overline{\cS}$ and any $s=(\alpha,x,\rho)\in\cS^*\backslash\overline{\cS}$, $u_i(\alpha^*)>u_i(\alpha),$ for all $i\in\cI$.
\end{definition}

Note that the set of payoff-dominant states may not be empty in a coordination game of Definition~\ref{def:CoordinationGame}, since the set of pure Nash equilibrium states is not empty. However, it might be the case that $\overline{\cS}\equiv\cS^*$.

\begin{proposition}	\label{Pr:StochasticStabilityCoordinationGames}
Consider a coordination game of Definition~\ref{def:CoordinationGame}, and let every agent implement the learning dynamics of Table~\ref{Tb:ReinforcementLearning} with sufficiently small $\epsilon>0$ such that $\epsilon u_i(\alpha) < 1$ for all $i$ and $\alpha\in\cA$. Then,
\begin{equation}
\lim_{h,\lambda\downarrow{0}}\mu_{\lambda}(\overline{\cS}) = 1.
\end{equation}
\end{proposition}
\ifproofs
\begin{proof}
(sketch) According to Theorem~\ref{Th:StochasticStability}, we have 
$\lim_{\lambda\downarrow{0}}\mu_{\lambda}(\overline{\cS}) = \hat{\mu}(\overline{\cS}).$ Thus, it is sufficient that we investigate the limit of $\hat{\mu}(\overline{\cS})$ as $h\downarrow{0}$. For any two pure Nash equilibrium states $s^*\in\overline{\cS}$ and $s\in\cS^*\backslash\overline{\cS}$, and for any $\delta>0$ 
$Q\Pi(s^*,\mathcal{N}_{\delta}(s))$ is scaled by $h$ (since the utility received under $s$ is strictly less than the utility received under $s^*$ for all agents). The conclusion follows by taking the limit as $h\downarrow{0}$.
\end{proof}
\fi

\begin{figure}[th!]	
\includegraphics[scale=1]{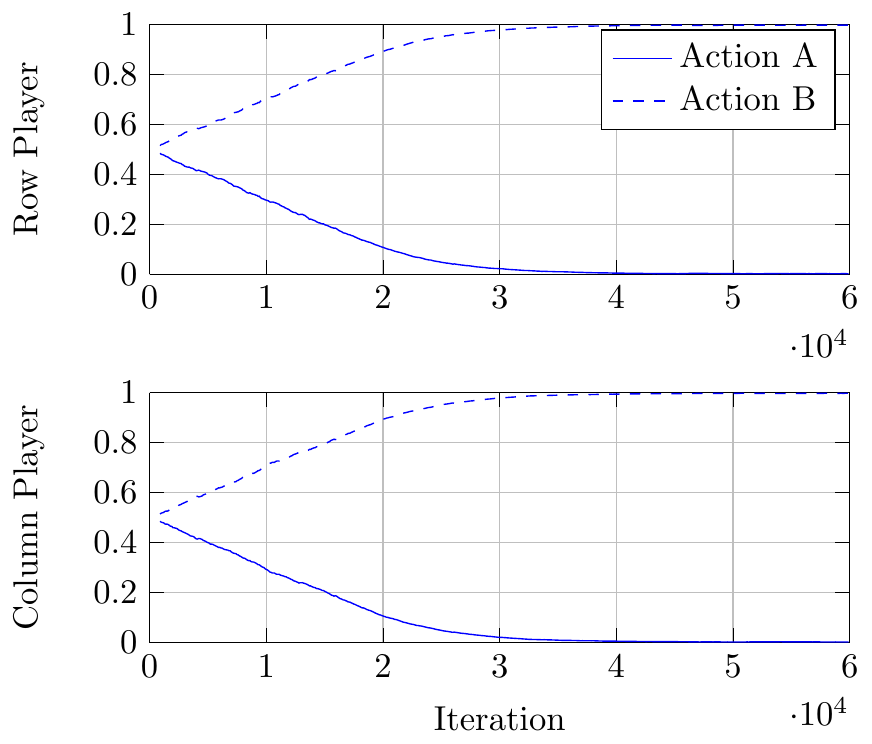}
\caption{Response of standard perturbed learning automata (PLA) ($h\equiv{1}$).}
\label{fig:SimulationStandard}
\end{figure}

\begin{figure}[th!]	
\includegraphics[scale=1]{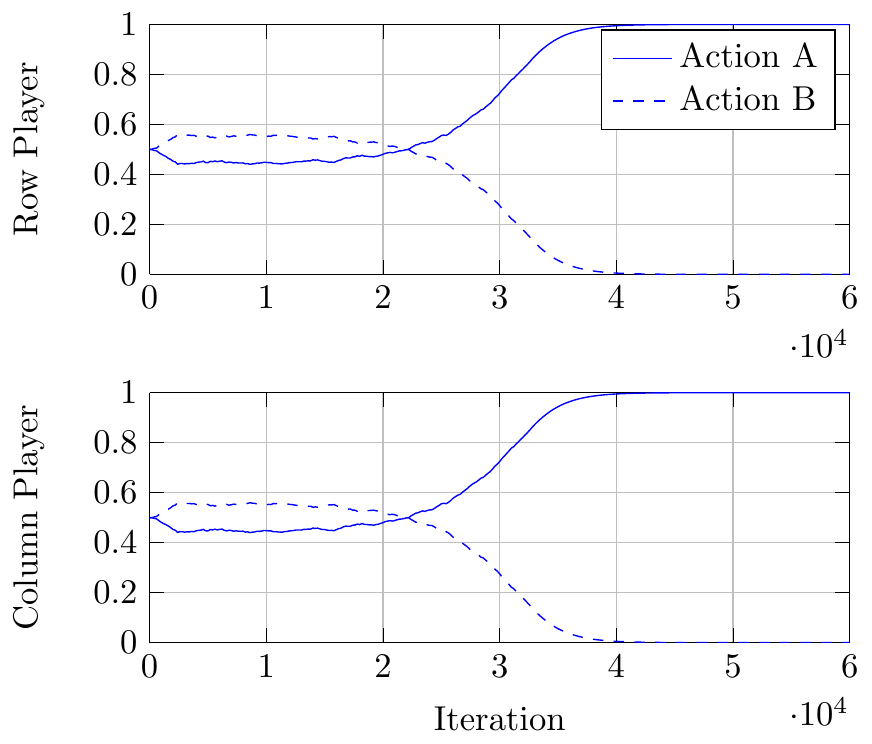}
\caption{Response of aspiration-based perturbed learning automata (APLA) (small $h>0$).}
\label{fig:SimulationAspirationBased}
\end{figure}

\subsection{Simulation study} \label{sec:SimulationStudy}

We consider the Stag-Hunt coordination game of Table~\ref{Tb:SHG} where agents implement the APLA dynamics of Table~\ref{Tb:ReinforcementLearning} with $\epsilon=0.0001$, $\nu=0.001$, $\lambda=0.01$ and $h=0.01$. The response of the aspiration-based reinforcement learning is shown in Figure~\ref{fig:SimulationAspirationBased}, while the corresponding response under standard perturbed learning automata (i.e., for $h\equiv{1}$), is shown in Figure~\ref{fig:SimulationStandard}. It is evident that under standard perturbed learning automata, the risk-dominant equilibrium, which corresponds to the action profile $(B,B)$ is the stochastically-stable state. On the other hand, Figure~\ref{fig:SimulationAspirationBased} demonstrates that the payoff-dominant equilibrium (which corresponds to the action profile $(A,A)$) is the stochastically-stable state under APLA, which verifies the conclusion of Proposition~\ref{Pr:StochasticStabilityCoordinationGames}.

\section{Conclusions}	\label{sec:Conclusions}

This paper introduced a novel learning automata dynamics, namely \emph{aspiration-based perturbed learning automata} (APLA). Contrary to standard perturbed learning automata, actions are reinforced based on both repeated selection and their satisfaction levels. This modification, that is solely based on local information available to each agent (i.e., its own utility function), is able to alter the stochastically stability properties of standard learning automata in strategic-form games. In particular, the payoff-dominant (pure) Nash equilibrium states become the only stochastically stable states. This overcomes several limitations observed in reinforcement-based learning dynamics (e.g., discrete-time replicator dynamics or learning-automata) where risk-dominant pure Nash equilibria can be the only stochastically stable outcomes. 


%
%
%
%

%
%
%
%
%
%

\bibliographystyle{IEEEtran}
\bibliography{2017_ECC18_AspirationBasedReinforcementLearning_Bibliography}

\end{document}